\documentclass[preprint]{aastex}
\def\asec{\ifmmode ^{\prime\prime}\else$^{\prime\prime}$\fi}
\def\etal{{et\,al. }}
\def\msun{M$_{\odot}$}

\def\degs{\ifmmode ^{\circ}\else$^{\circ}$\fi}
\def\amin{\ifmmode ^{\prime}\else$^{\prime}$\fi}
\def\asec{\ifmmode ^{\prime\prime}\else$^{\prime\prime}$\fi}
\def\fm{\hbox{$.\!\!^{\rm m}$}}            % Fractions of magnitudes
\def\fdg{\hbox{$.\!\!^\circ$}}          % Fractions of degrees
\def\fhour{\hbox{$.\!\!^{\rm h}$}} % Fractions of hours

\def\degs{\ifmmode ^{\circ}\else$^{\circ}$\fi}
\def\amin{\ifmmode ^{\prime}\else$^{\prime}$\fi}

\begin{document}
\title{FS Aur - a new class of Cataclysmic Variables \\ or the  missing link between Intermediate Polars 
and SW Sex objects? }

 \author{Gaghik Tovmassian\altaffilmark{1},
         Sergei Zharikov\altaffilmark{1},
         Raul Michel\altaffilmark{1}}
\affil{Observatorio Astron\'omico Nacional, Instituto de
              Astronom\'\i a, UNAM, P.O. Box 439027, San Diego,
              CA, USA 92143-9027 }
              \email{\{gag, zhar, rmm\}@astrosen.unam.mx}
\and
 \author{Vitaly Neustroev\altaffilmark{2}}
\affil{Department of Astronomy and Mechanics, Udmurtia State University, 
              Universitetskaia 1, Izhevsk 426034, Russia}
              \email{benj@udm.ru}
\and
 \author{Jochen Greiner\altaffilmark{3}}
\affil{Max-Planck-Institut f\"{u}r extraterrestrische Physik, Giessenbachstrasse
Postfach 1312, 85741 Garching, Germany} 
\email{jcg@mpe.mpg.de}
\and
 \author{David R. Skillman\altaffilmark{4}}
 \affil{Center for Backyard Astrophysics (East), 9517 Washington Avenue, Laurel, MD
20723}
\email{dskillman@home.com}
\and
\author{David A. Harvey\altaffilmark{5}}
\affil{ Center for Backyard Astrophysics (West), 1552 West Chapala Drive, Tucson, AZ
85704}
\email{dharvey@comsoft-telescope.com}
\and
\author{Robert E. Fried\altaffilmark{6}}
\affil{Center for Backyard Astrophysics (Flagstaff), Braeside Observatory, P.O. Box 906,
Flagstaff, AZ 86002} \email{captain@asu.edu} 
\and
\author{Joseph Patterson\altaffilmark{7}}
\affil{Department of Astronomy, Columbia University, 550 West 120th Street, New
York, NY 10027}
\email{jop@astro.columbia.edu}

\begin{abstract}
        FS Aur is  a known dwarf nova with an  orbital period of about
        85.7 minutes.   It has  been assumed  to be a  member of the SU UMa
        subclass  of  cataclysmic variables (CVs), but  previous  
        searches  for superhumps  and
        superoutburst  have been unsuccessful.  We  conducted a
        series of  photometric and spectroscopic  observations of FS Aur
        during  quiescence.  We  confirmed its  short  orbital
        period  from radial velocity measurements.   
        However, the long-term  photometry
        revealed  an  unexpected  result:  the  system  also  shows  a
        distinct 0.24  mag modulation in the
        BVR photometric bands with  a period of 205.5 min, which
        is  2.4  times longer  than  the  orbital  period. We  discuss
        various possible causes for such a peculiar behavior.
\end{abstract}  


%________________________________________________________________

\section{Introduction}

FS  Aur was discovered  back in  1949 \cite{hoff49}  as a  dwarf nova.
\cite{How88} made  an attempt to  determine its orbital  period, among
several  other  systems, by  means  of  CCD  photometry. They  cite  a
spectroscopic period of  85 min, but they failed  to find any definite
period  from   photometry,  although  they  have   detected  0.15  mag
variations   which  could   be   consistent  with   87   to  105   min
periods.  Later,   \cite{Thor96}  refined  the   orbital  period  from
spectroscopy, and suggested a SU UMa  classification 
based on the short orbital period of 0.0595 days (85.7 min).

   Nevertheless, long-term monitoring of the system by several groups failed
to  detect  any  superoutburst/superhumps  in  its  light  curve.  The
publicly available  VSOLJ and AAVSO light curves  show instead a steady-cyclic
outburst pattern  that  is more similar to a SS Cyg  type dwarf
nova lightcurve. \cite{And91} found 12 outbursts
on 180 archival photographic plates of the Sternberg State Astronomical
Institute.  But he also failed to  find any super-outbursts in the archival
data and notes the regular character of the outbursts, similar to dwarf novae.

   \cite{Neu02},  hereafter   N02,  among  others, for a long time hunted  for
superoutbursts and  also conducted  a new study  of the system  in 
quiescence.  In  his  paper  he  noted  several  interesting  spectral
characteristics  of the object, roughly determined the system parameters
and  also  discovered  variability  with a period longer  than the  
orbital  period.

   Here we present the results of a new long-term photometric campaign and
additional spectroscopic observations  of FS  Aur (\S 2). They  are combined
with  data previously  obtained  by N02,  separated  by a  3 year  time
gap. We compare the properties of FS Aur with SW Sex stars (\S 3.2), and
describe our discovery of the existence  of a 3.42 hour photometric period in
the light curve of FS Aur (\S 3.3).  Finally, we discuss possible causes of 
this unusual
behavior of the system (\S 4) and possible implications on the classification
of Cataclysmic Variables(\S 5) .

\section{Observations}

\subsection{Optical Observations}

New observations  of FS Aur were obtained  at the Observatorio Astronomico
Nacional  (OAN  SPM)  in  Mexico.   The 2.1\,m  telescope  with the B\&Ch
spectrograph,  equipped with a SITe $1024\times1024$ pixel CCD with a pixel
size of  24\ $\mu\rm m$, was used  for the spectroscopic  observations.  
Spectra were obtained in the second  order of a 400  l/mm grating, 
covering the 3800--5200\AA\, region
and reaching 2.5 \AA\,  FWHM spectral resolution with the used 
slit width of 1.5\asec. Spectral observations
were done in  two consecutive nights (31 Jan/1 Feb  2001) with a total
coverage of ~12 hours, consisting of  10 min duration individual exposures. 
He-Ar arc lamp
exposures  were  taken  before   and  after the object  observations  for
wavelength  calibration, and the standard spectrophotometric  star G191B2B
\citep{Oke90}    was    observed    each   night    for    flux
calibration. The reduction  of the images and the extraction of the spectra  
was done using standard 
IRAF\footnote{IRAF is distributed by the National Optical Astronomical
Observatories, which  is operated  by the Associated  Universities for
Research in  Astronomy, Inc., under  contract to the  National Science
Foundation.} routines customly used for long-slit spectra.

   Photometric observations were performed with the 1.5\,m telescope of OAN
SPM, and  cover a longer  time span.  Differential photometry  was obtained
with the three  broad  band   Johnson-Cousins  filters  B\,V\,R  in  long
time-series  with individual  exposures ranging  from 20  to  180 sec.
Selected  comparison and  check  star  form the  field  were used  for
differential  photometry   performed  using the DAOPHOT   package  within
IRAF.

Additional  photometry was acquired  by members  of the Center  of Backyard
Astronomy (CBA)\footnote{http://cba.phys.columbia.edu/}. 
For details of  their activity and procedures used to
obtain and reduce the data we refer to \cite{Patt00}.

The log of all optical observations is presented in Table \ref{tab1}.

\subsection{ROSAT X-Ray Observations}

We also retrieved all FS Aur X-ray observations from the ROSAT archive
at MPE Garching. FS Aur was observed in the course of the 
all-sky survey in August/September 1990 for a total exposure time of 441 sec,
and is listed in the ROSAT All-Sky Survey Bright Source Catalog as
1RXS J054748.5+283512. It is detected at a count rate of 0.13$\pm$0.02 cts/s.

FS  Aur  was  serendipitously  observed  again  in  pointed  mode  on
23 September 1993  with  the position-sensitive proportional counter 
(PSPC, Pfeffermann \etal\ 1986)
with  a total  exposure time of 9509\,sec. Due to its off-axis angle,
the effective exposure of FS Aur is 6130\,sec.
The mean count rate is 0.128$\pm$0.005 cts/s, identical to that of
the all-sky survey observation in 1990.

The spectrum is identical in shape and brightness during both
observations.  The hardness  ratio  of the  object  inferred from  these
observations is  HR1 = 0.911$\pm$0.027  and HR2 =  0.371$\pm$0.034,
These hardness ratios are defined as HR1 = $(B-A)/(B+A)$ and 
HR2 = $(D-C)/(D+C)$,
 where $A (0.1-0.4$ keV), $B (0.5-2.0$ keV), $C (0.5-0.9$ keV),
 and $D (0.9-2.0$ keV) are the counts in the given energy range.
 Given the  galactic coordinates of FS Aur of 
L{\sc$_{II}$} =  180\fdg55, B{\sc$_{II}$}  = +0\fdg23,
these  hardness ratios  can be  interpreted  as a  rather hard  source
spectrum, with  strong interstellar absorption which practically leaves 
no counts below 0.5 keV.

\begin{figure}[]
\plotone{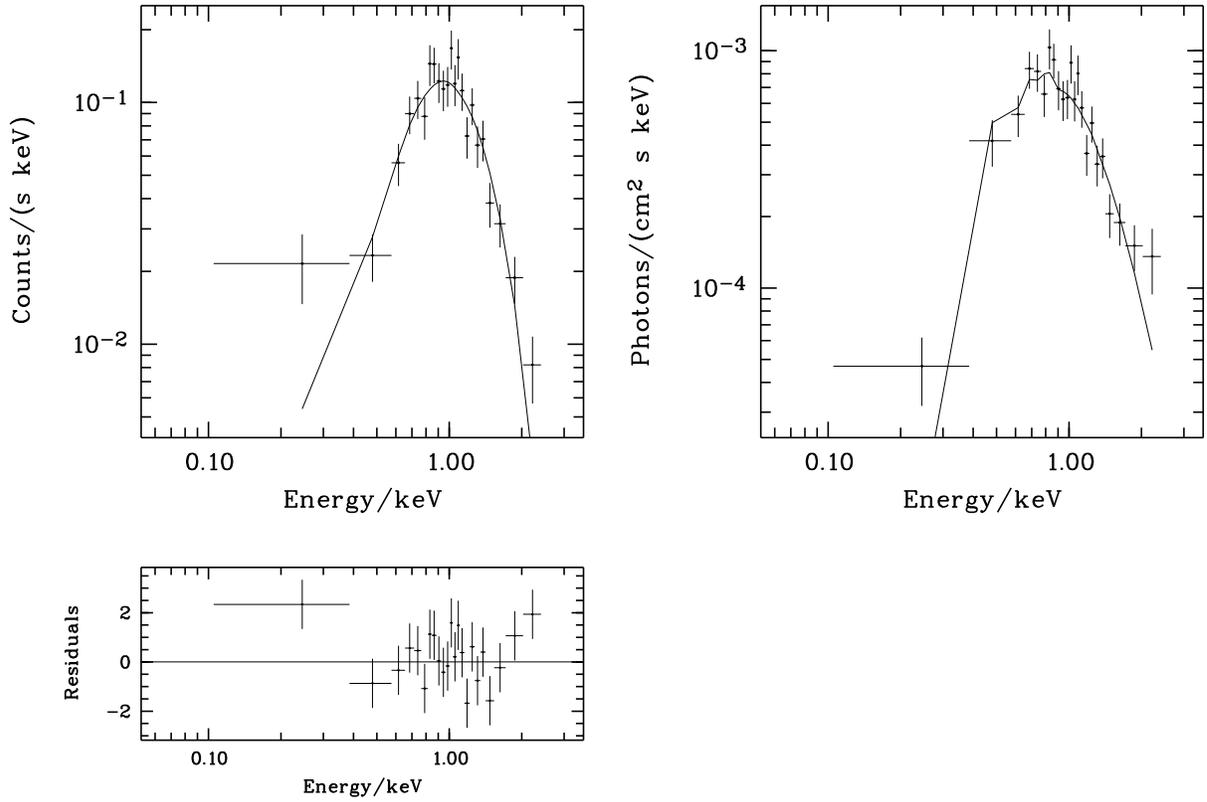}
\caption{ ROSAT  PSPC X-ray spectrum  of FS Aur  derived from
the pointed observation from 23 September 1993, fitted with a
thermal bremsstrahlung model (for fit parameters see text). 
The lower left panel shows the deviation
between data and model in units of $\chi^2$.
}
\label{xspec}
\end{figure}

Due to this strong absorption and the restricted spectral coverage, 
several spectral models fit the data.
The spectral fit  using a  thermal bremsstrahlung model is
presented on Figure \ref{xspec}, and gives the following parameters:
absorbing column $N_{\rm H}$ = (3.0$\pm$0.3)$\times$10$^{21}$ cm$^{-2}$, 
temperature  $kT$ = 0.59$\pm$0.30 keV, and a normalization of
(1.3$\pm$0.1)$\times$10$^{-3}$ ph/cm$^2$/s/keV.
The absorbing column corresponds to about half the total galactic column
in this direction (Dickey \& Lockman 1990).
The unabsorbed bolometric flux for this model is 
1.0$\times$10$^{-11}$ erg/cm$^2$/s, and the luminosity is 
1.2$\times$10$^{31}$ (D/100 pc)$^2$ erg/s.

\section{Results}

\subsection{Spectroscopy}

Our spectroscopic data basically repeat all characteristics described
by  N02. For  radial  velocity  (RV) measurements  we  simply did  one
gaussian fitting  to the mostly  single peaked H$_\beta$  emission line
profiles.  Using RV measurements from our observations and combining them with the
measurements from N02, we derived  the spectroscopic (orbital) period of 
FS Aur.  We used the DFT method  to  obtain the power  spectrum, the  spectral
window   corresponding  to the  data   and the time   distribution  of   our
observations.  
%The  power spectrum  of  the  H$_\beta$ radial  velocity
%measurements from both data sets  is presented in Figure \ref{powsp}. The
%overplotted thick curve  is the CLEANed power spectrum of  the same data set.
The period analysis resulted in a strong peak confirming previous determination
of the orbital period. Using the CLEAN procedure \citep{Roberts} that allows to 
suppress alias frequency peaks caused by uneven data sampling, we obtained 
the strongest  peak  at  the  frequency corresponding to 85.65 (0\fd059479)  
minutes. 
This estimate is within the range of periods and margins of error cited by previous 
authors  \citep{Thor96}. 
There  is another peak  at around 10
day$^{-1}$.  We will  discuss  this later in the text after considering our 
photometric  data 
(\S 4).  The superior conjunction of the
emission  line  source (i.e.  red-blue  crossing)  corresponds to  HJD
T$_0$=2450101.2656.  However,  since the  source  of  emission is  not
defined, it is not clear whether it is a superior conjunction of the WD
or not.

Other emission lines show 
similar results with no significant phase shifts between the lines.

\begin{figure}[ht]
\plotone{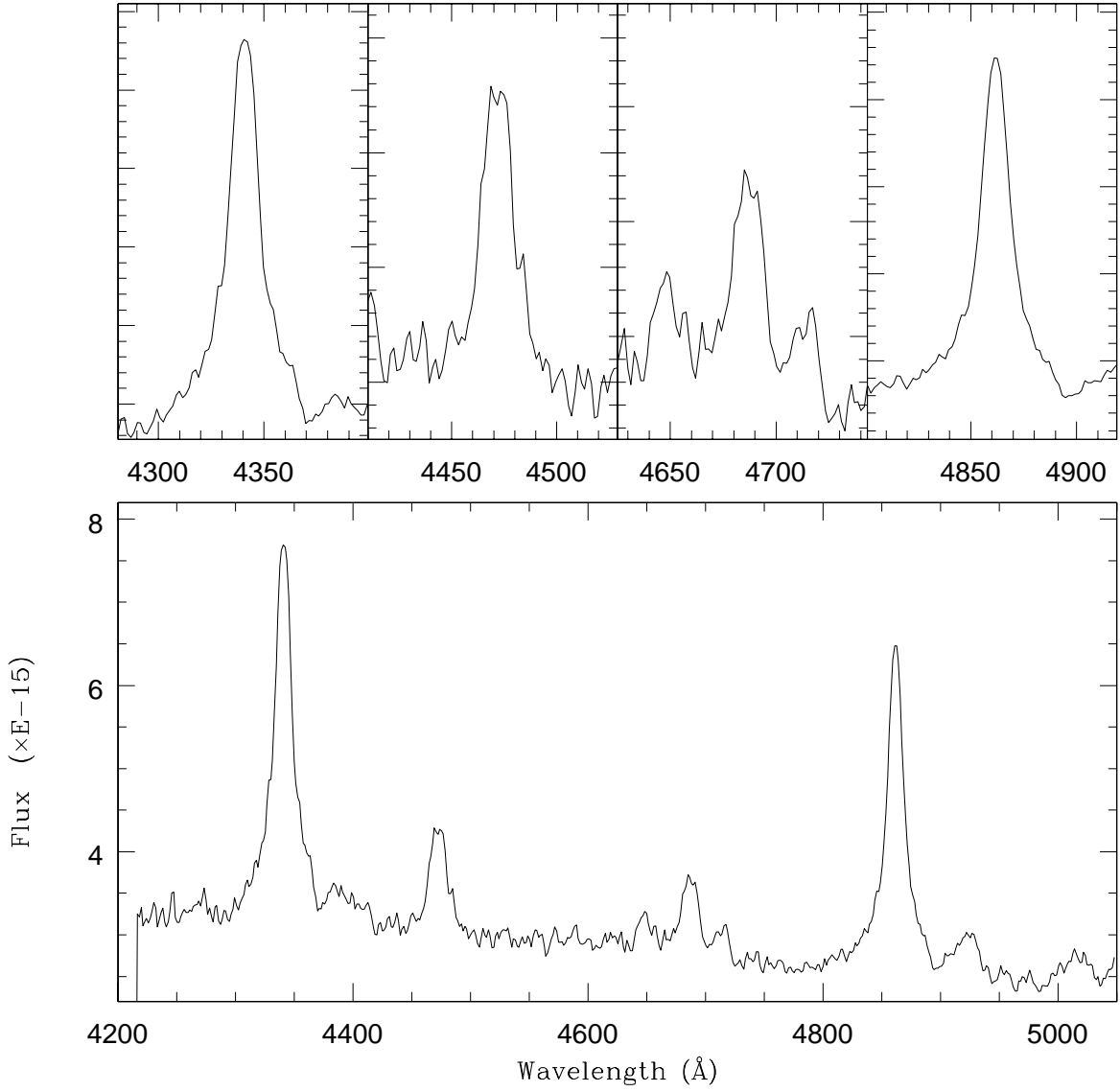}
\caption{Averaged spectrum  of FS  Aur, obtained  by  co-adding all  obtained 
spectra after  correcting for
orbital  radial  velocities  changes.
}
\label{spectrum}
\end{figure}

N02  estimates the  inclination  angle  of   the  system  to   be  within
51\degs-65\degs. While it is based on dynamical interpretation of emission 
lines which might be highly  unreliable indicator of the motions of the underlying stars, 
it  is in a  good agreement with the velocities obtained
from our new observations. It is surprising though not to see broad double
peaked   emission  lines   from this short-period system (Dwarf   Nova)
at this inclination. It is noteworthy, that  He\,{\sc ii} $\lambda4686$ \AA\, shows up strongly. 
It is unusual to have He\,{\sc ii} this strong in an ordinary dwarf nova.
It appears that the emission lines  of higher excitation
show central  absorption cores at  some phases while  H$_\beta$ remains
mostly  single-peaked.   Probably because there is  no  systematics behind
this behaviour, we
could  not find  any firm  phase dependent  pattern.  We  averaged the
spectrum  of FS  Aur by  co-adding all  spectra after correcting for the
orbital  radial  velocity  (Figure
\ref{spectrum}). Individual lines are blown up in  separate panels
to   display the  abovementioned   features.   These   and   some   other
characteristics noted in FS Aur bring us to the idea that it may
belong to the class of SW Sex objects.

\subsection{Is FS Aur an  SW Sex object?}

  Careful examination of spectra of FS Aur shows that:
\begin{itemize}

\item Inspite of the estimate of a relatively high  inclination angle of
the binary  system of  $51 <  i < 65$,  emission lines  of FS  Aur are
single peaked. However, distorted wings can be seen in form of blue shifted
shoulders around phase 0.8 (N02).

\item  Central small  absorption can  be seen  in the higher  Balmer lines
starting  from H$_\gamma$,  and  in the He\,{\sc i}  4471  and He\,{\sc ii}  lines at  some
phases.  The  absorption  core of  He  lines  could  be seen  even  in the
integrated spectrum (see Figure \ref{spectrum}).

\item Although there are no eclipses, high excitation lines do notably
diminish at some phases (N02).
\end{itemize}

\begin{figure*}[ht]
\plotone{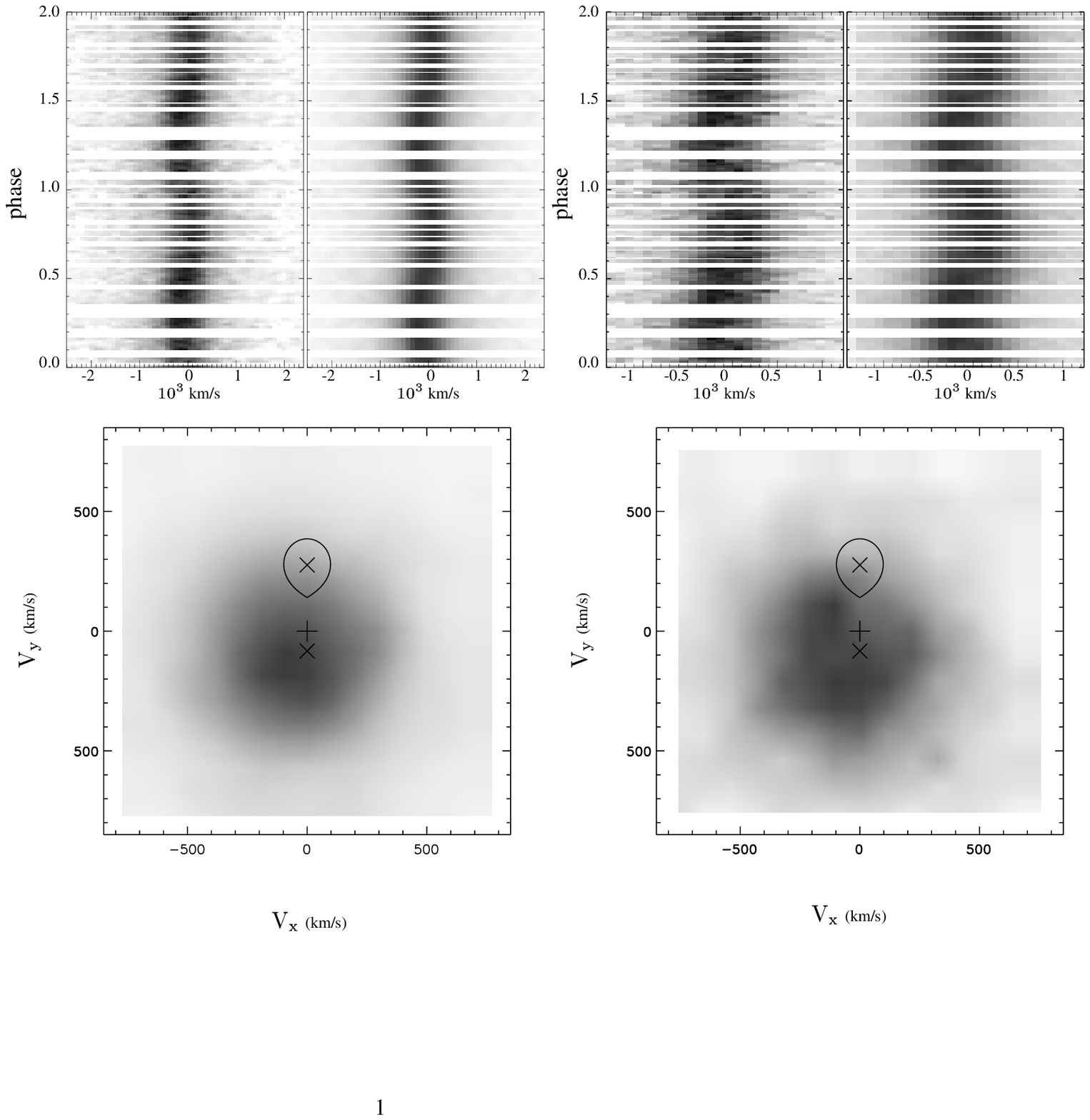}
\caption{ The velocity  maps, trailed and
reconstructed   spectra  of   H$_\beta$   and  H$_\gamma$. 
}
\label{tomo}
\end{figure*}

  These  features are  proper to  SW Sex  objects \citep{Tor91,Hel00}.
Although most  of the  SW Sex stars  are eclipsing  systems (having been
one of the original class identifiers), recently also
low-inclination systems were discovered  pertaining to that class.  In
particular, FS Aur shows remarkable similarities in the line profiles to LS
Peg  \citep{Tay99} and V442  Oph \citep{Hoa00}.   The shoulder  in the
blue wing of H$_\beta$ around phase 0.8 is similar to those observed in
LS Peg  and V442 Oph,  where it is  identified with the  high velocity
component  in these systems.  We were  not able  to separate  the high
velocity component, probably because of  the low S/N ratio of our data.
Also worth  mentioning is the relatively  small intensity of the He\,{\sc ii}  line in
both  these  low-inclination  systems.   Finally,  there  is a  striking
resemblance of  the Doppler tomograms of  FS Aur with those  of SW Sex
stars.  The ring in  the maps,  corresponding to  the accretion  disk, is
rarely  seen  in SW  Sex  systems.  Instead,  bright spots  are  often
detected in  the lower left quadrant  or below the location  of the WD,
and they  may vary from line  to line. The velocity  maps, the trailed and
reconstructed   spectra  of   H$_\beta$   and  H$_\gamma$ of FS Aur  are   
presented  in   Figure
\ref{tomo}. Similar  tomograms were obtained by  Neustroev (N02). Since
the phasing  in the absence of  eclipses or any other  clear indication of
zero phase is arbitrary, the location  of the spots on velocity map is
arbitrary too.

Nevertheless, it should be noted that other characteristics of FS Aur do not 
fit well into the SW Sex definition. The latter are mostly novalike systems 
exhibiting high and low states of luminosity, while FS Aur undergoes 
$\approx2$ mag dwarf nova like outbursts.
Another significant difference between FS Aur and the SW Sex stars is its
short orbital period.
However, the spectral (orbital) period, unambiguously detected in FS Aur,
appears to be in large contrast with the  photometric period discovered by us.

\subsection{Photometric Period}

\begin{figure}[ht]
\plotone{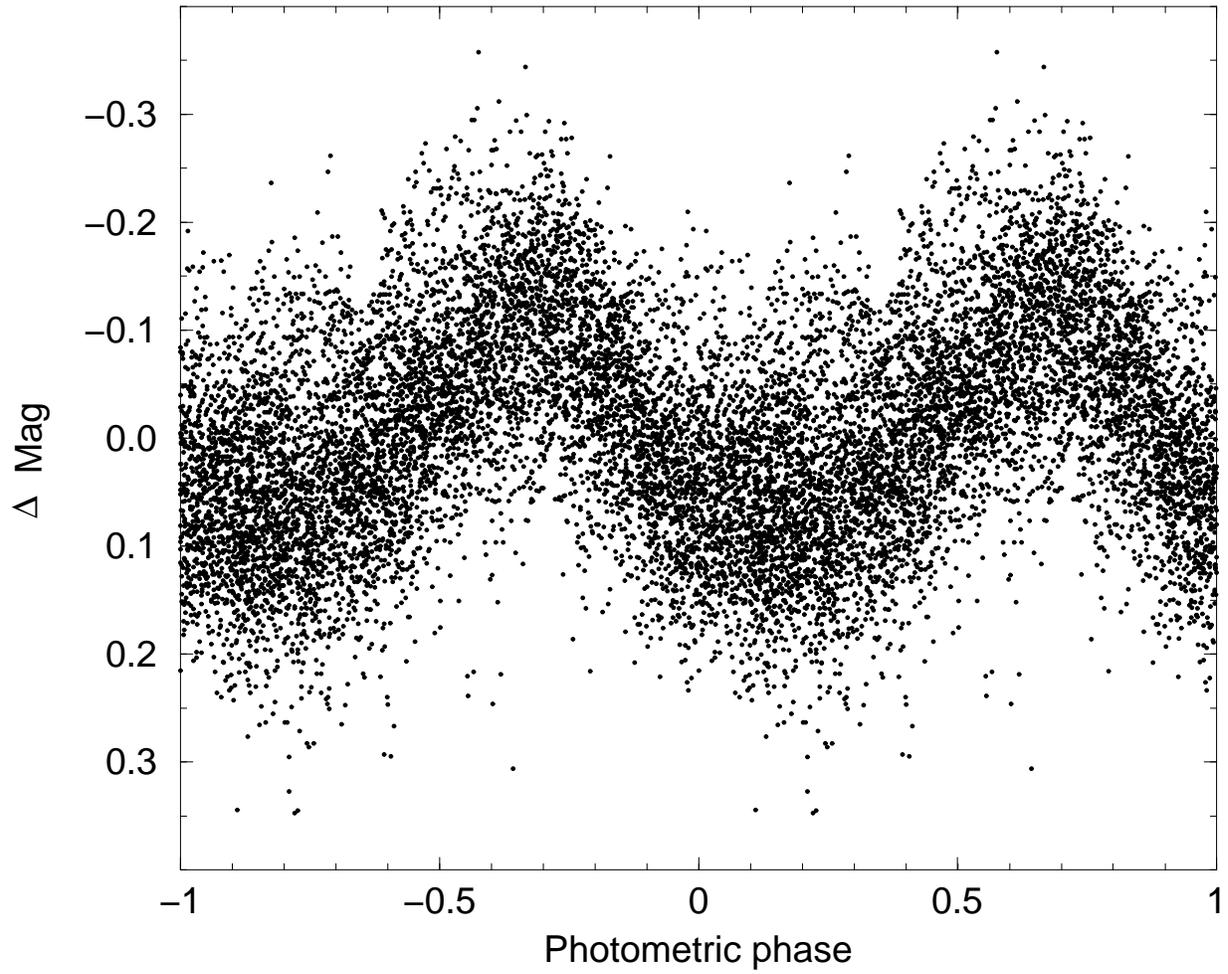}
\caption{Several thosands of photometric measurements collected from a
variety of CBA
small telescopes  in addition to the SPM observations  reported earlier in
\cite{Tov01},  folded  with  the  photometric  period.  The  data  were
normalized  to each night's  mean magnitude,  since they  were obtained
with various  detectors and comparison  stars.
}
\label{comblc}
\end{figure}

The   extensive  photometric  coverage   reveals  a   most  unexpected
result. The long-term variability  noted in N02 in reality  appears to be
strictly periodic.  All data that we  were able to  collect since then
show the same period of  3\fhour4 and easily can be phase-folded over
the time exceeding 3 years  of observations. In Figure \ref{comblc} we
present several thousands of measurements collected from a variety of CBA
small telescopes,  in addition to SPM observations  reported earlier in
\cite{Tov01} and folded  with  the  photometric  period.  The  data  were
normalized  to each night's  mean magnitude,  since they  were obtained
with various  detectors and comparison  stars.  In spite of  the large
0\fm2  scatter   the  sinusoidal  shape  of the light   curve  is  mostly
undisturbed.  Neither the phase nor the amplitude of  variations seem  to be
altered  by the cyclic  outbursts  that the system  underwent  in between  the
different sets  of observations.  This demonstrates the  very coherent
nature of the periodic signal in the light curves.

\begin{figure}[ht]
\plotone{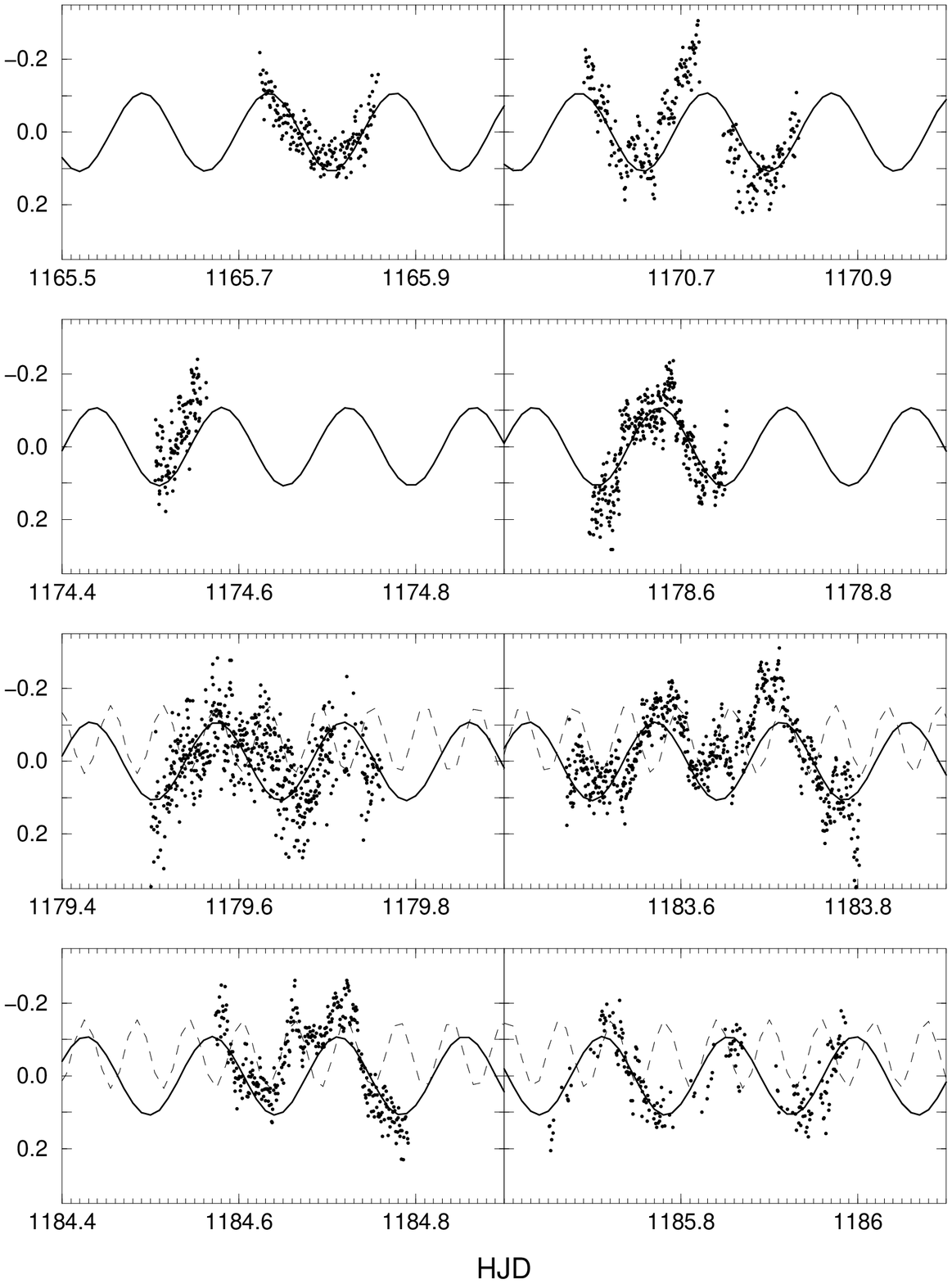}
\caption{FS Aur light curves are presented. The epoch of observations is 2450000 + HJD. Overlayed solid curve represents the photometrical period.
The dashed line corresponds to the  spectroscopic period and described in the text.
 }
\label{En1}
\end{figure}

\begin{figure}[ht]
\plotone{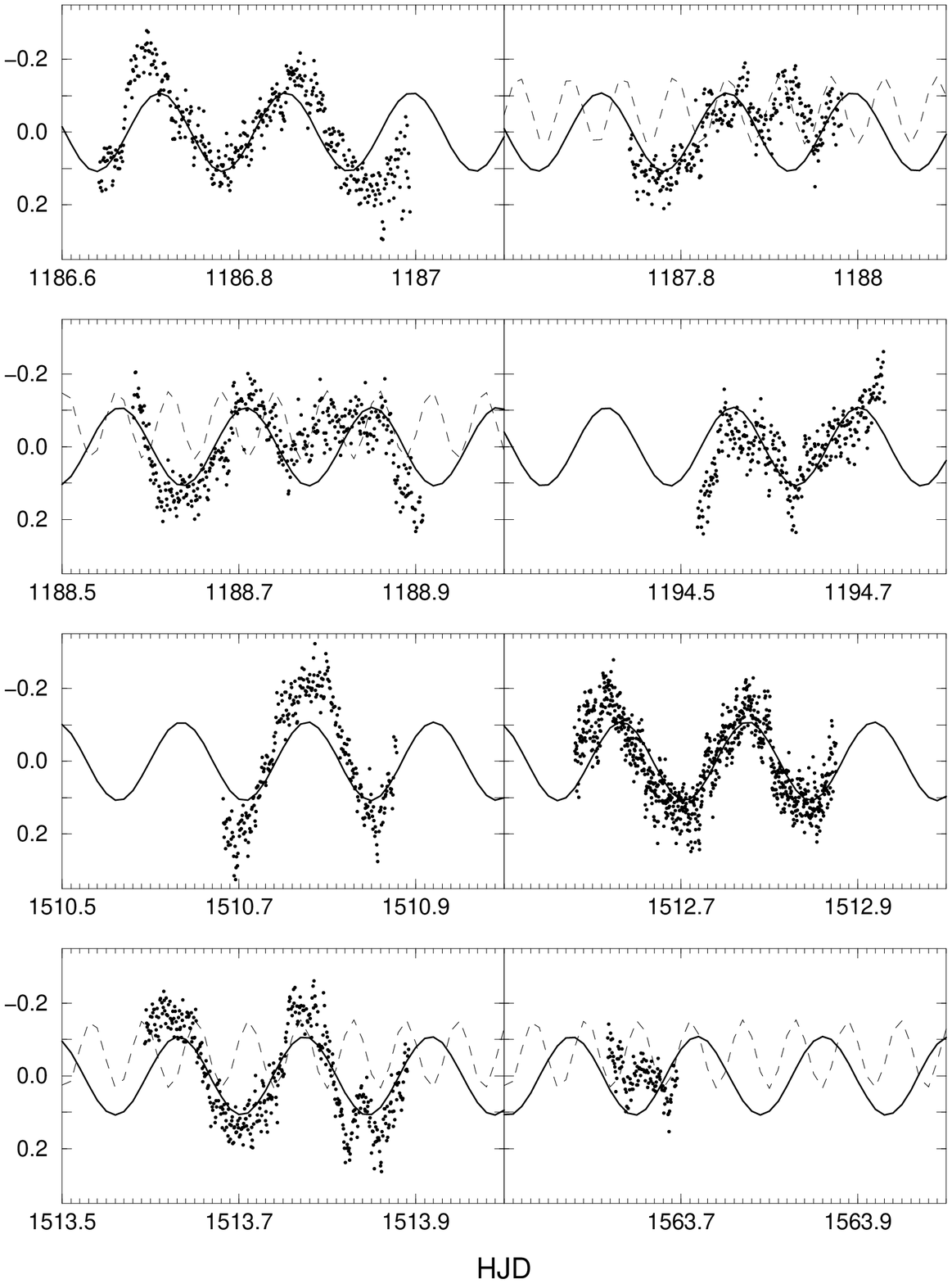}
\caption{FS Aur light curves are presented. The epoch of observations is 2450000 + HJD. Overlayed solid curve represents the photometrical period.
The dashed line corresponds to the  spectroscopic period and described in the text.
 }
\label{En2}
\end{figure}

\begin{figure}[ht]
\plotone{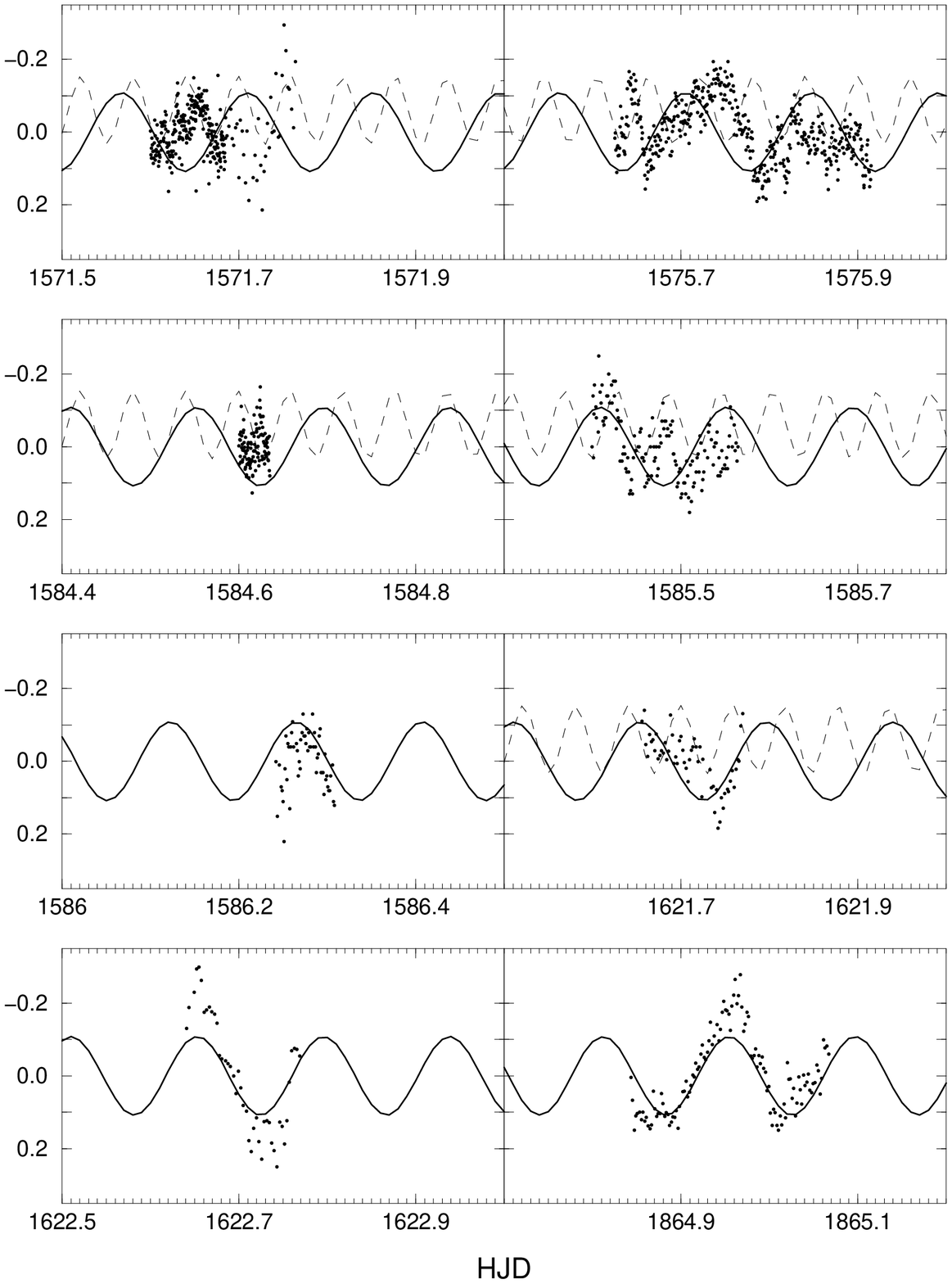}
\caption{FS Aur light curves are presented. The epoch of observations is 2450000 + HJD.  Overlayed solid curve represents the photometrical period.
The dashed line corresponds to the  spectroscopic period and described in the text.
 }
\label{En3}

\end{figure}

\begin{figure}[ht]
\plotone{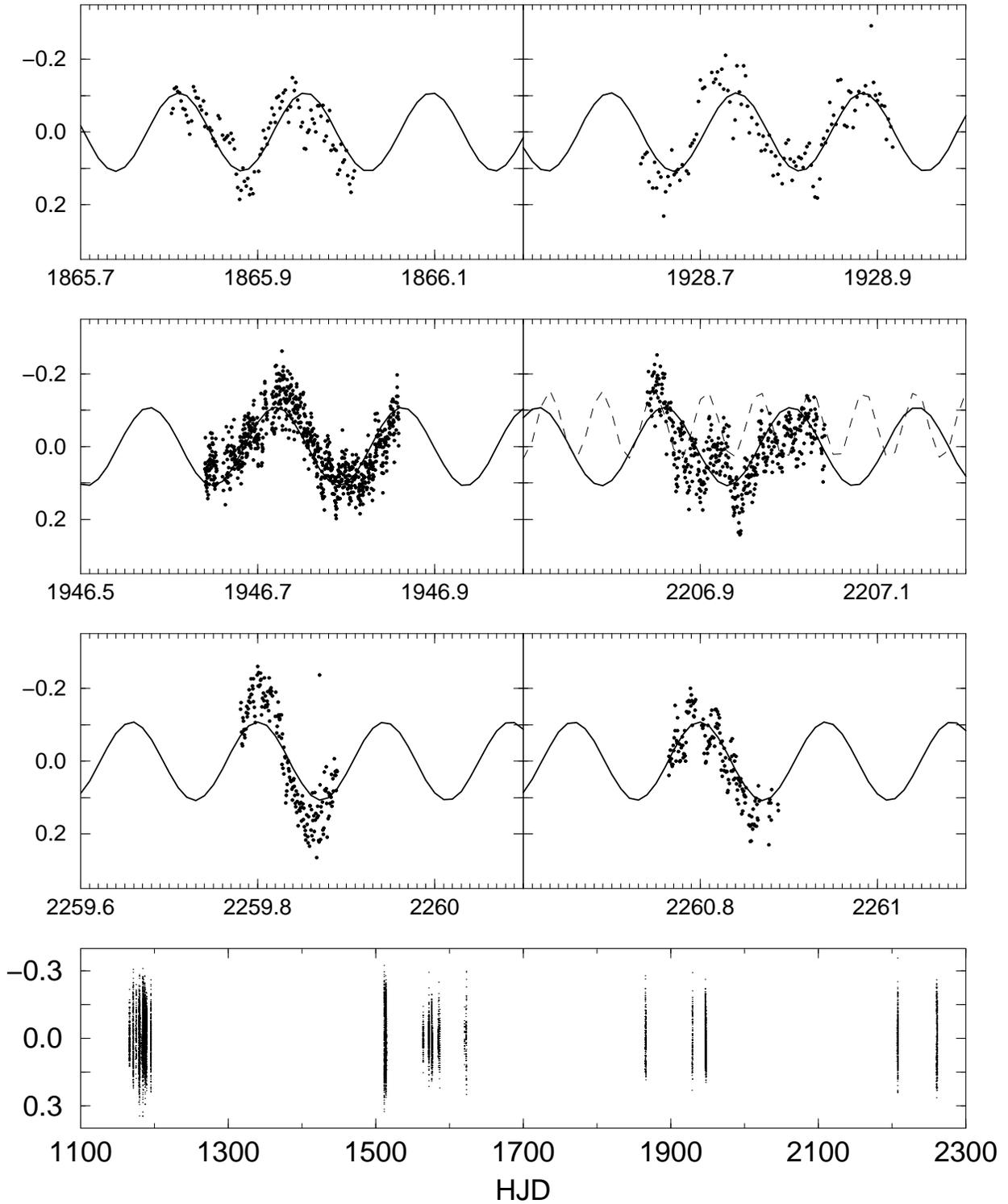}
\caption{FS Aur light curves are presented. The epoch of observations is 2450000 + HJD. Overlayed solid curve represents the photometrical period.
The dashed line corresponds to the  spectroscopic period and described in the text. The low panel shows   all data together.
 }
\label{En4}
\end{figure}

However, when  we look closely at each  night's data, we can  see that the
above mentioned periodic signal in the light curve varies in its strength
and appearance. Since the observed phenomenon is of unusual nature we
decided to present them all for clarity. In Figures \ref{En1}, \ref{En2},
\ref{En3}, \ref{En4}  each
night's  data is presented  in its  corresponding panel.  The overlayed solid
curve  represents  the photometric   period.  At  nights when  the
photometric period is not seen  clearly or there are faster variations
present, we have marked also the  spectroscopic period with a dashed  line.  
Its phase was estimated by a least-square fitting of the curve to the moment
of minima  and maxima measured at  these nights. We are  not sure that
these rapid variations are the result  of orbital modulations and these 
curves are presented solely for demonstration purposes.

\begin{figure}[ht]
\plotone{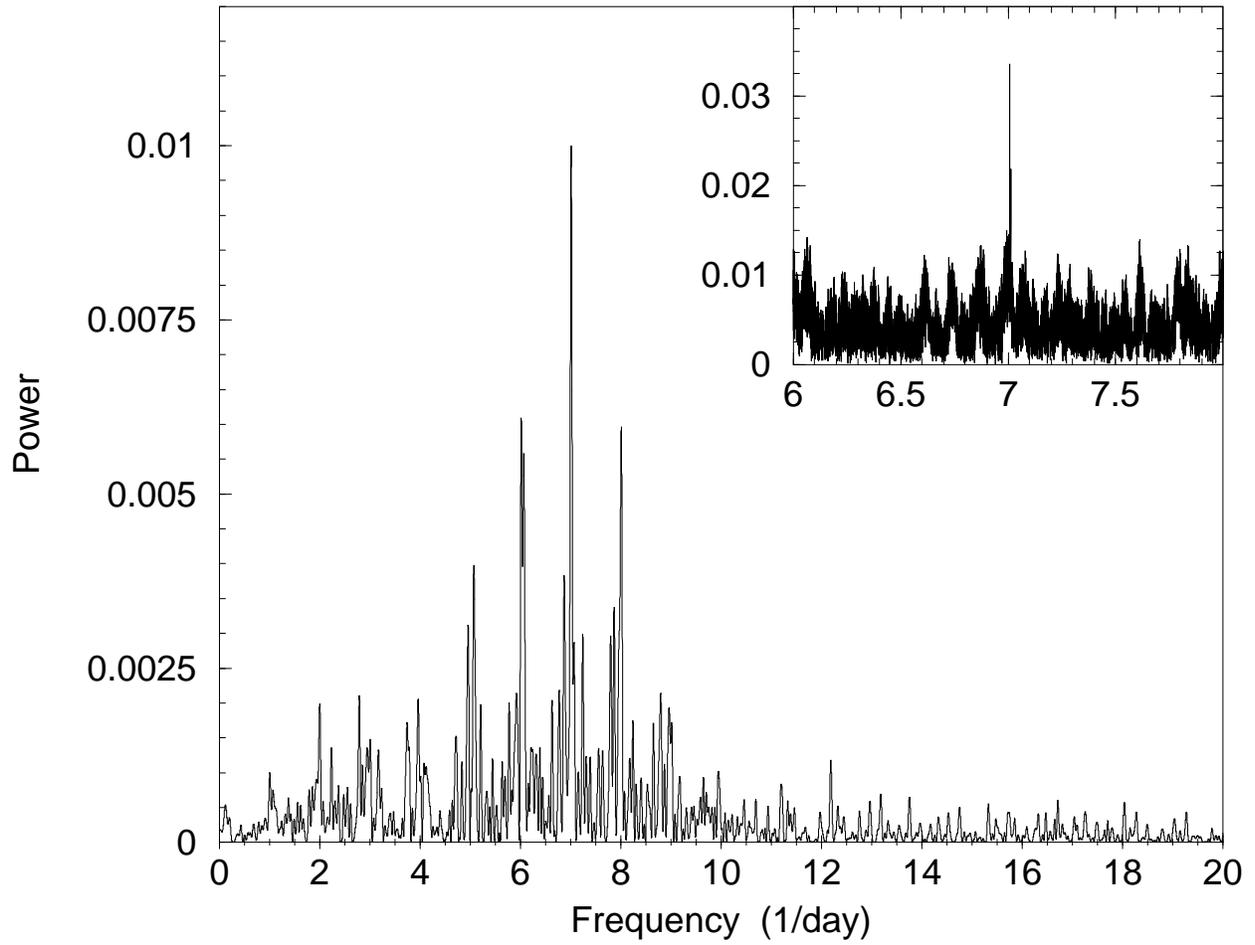}
\caption{The CLEANed  power spectrum of the combined photometric data and blowout of the 
       most prominent peak.}
\label{photpower}
\end{figure}

\begin{figure}[ht]
\plotone{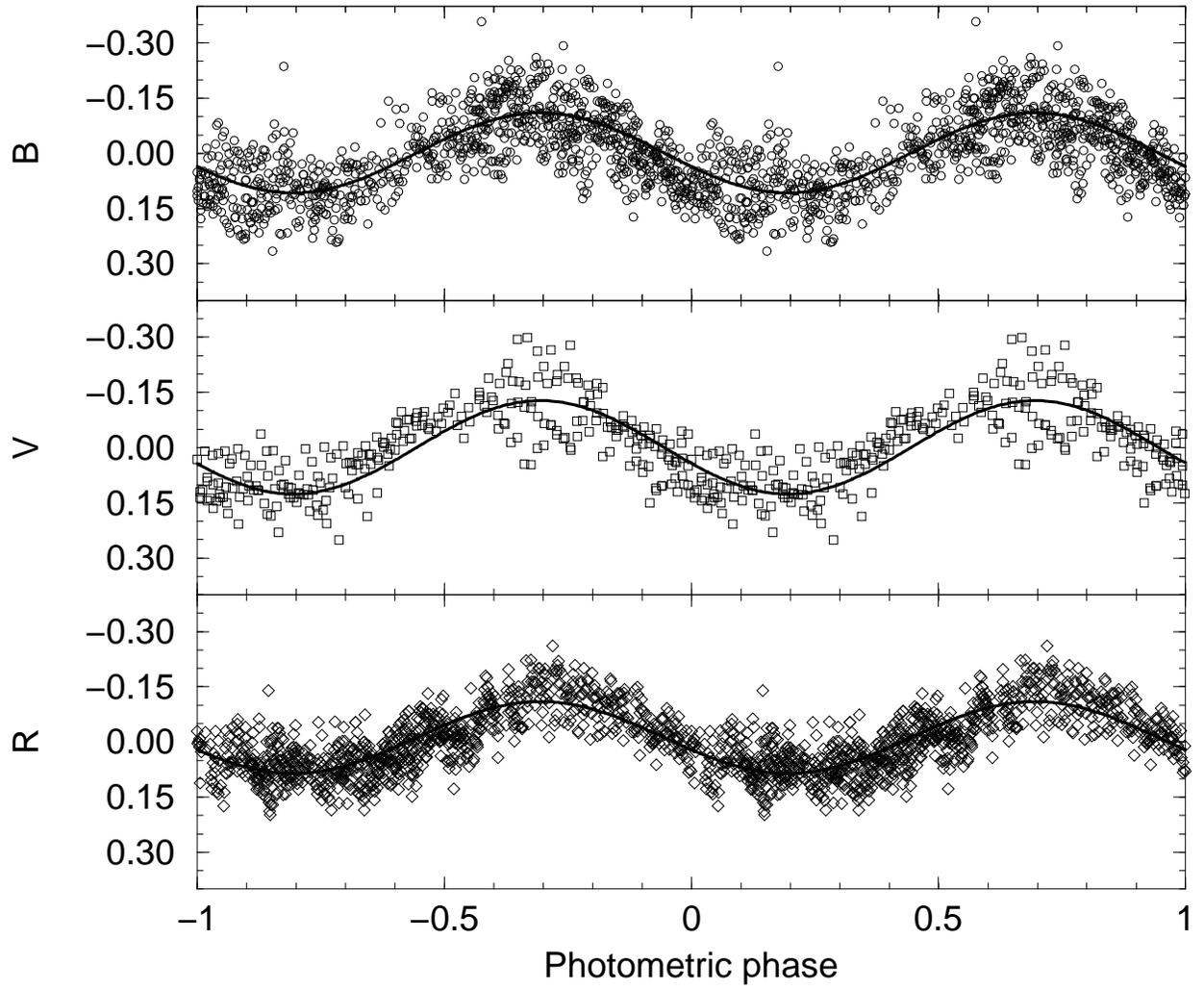}
\caption{BVR light curves of FS Aur folded with the photometrical period
205.5min.
 }
\label{bvrlc}
\end{figure}

In the broad-band filters obtained at the 1.5\,m telescope of OAN SPM,
the  system shows  smooth sinusoidal  light curves. The  data collected
from a variety of CBA small  telescopes is not as smooth, and additionally
spontaneous  wiggles are detected  in the  light curves  of individual
nights.  However,  the  3.4   hours  periodic  signal  can  be  clearly
distinguished  in the light  curves and  unambiguously stands-up  in the
power  spectrum.   The most  important  characteristics  of the  light
curves are:

\begin{itemize}
\item  The period of variation in the light curve is 
3\fhour4247$\pm$0\fhour0005  as estimated from the combined data 
(differential magnitudes of all dates 
       and  colors were mean-subtracted and combined in one large dataset). 
       The power spectra of individual nights were also calculated and 
checked for consistency.  Depending on the duration and data quality they 
all show a similar pattern within errors between individual nights. The 
CLEANed  power spectrum of the combined data and a blowout of the 
       most prominent peak  are presented on Figure \ref{photpower}.
\item  The B\,V\,R light curves from OAN SPM are smooth and almost perfectly  
sinusoidal (see Figure \ref{bvrlc}).
\item The amplitude of  0.24 mag is approximately equal in all three bands 
(B\,V\,R).
 
\item The individual light curves from the unfiltered CBA observations are
more ``noisy''.   By this we do  not mean photon statistics  or a data
scatter due  to the flickering which is usual  for CVs. Instead, the light 
curves just show
more erratic wiggling  imposed on a longer term  variation due to the
3\fhour4  period.  However,  the   period  is  coherent  over  about  3
years. The phase-folded  light curve of all combined data,
 presented  in Figure \ref{comblc},  clearly demonstrates this coherence.

\item  The examination of other peaks in the power spectrum of the 
radial velocities shows    
    an unidentified peak  
   at 9.88 day$^{-1}$, which is very close to the  side-band frequency of
\begin{equation} 
% \tau_{\mathrm{co}} = \frac{E_{\mathrm{th}}}{L_{r0}} \,,
    \frac{1}{\mathrm P_{sb}} = \frac{1}{\mathrm P_{orb}} - \frac{1}{\mathrm P_{phot}} 
\end{equation}  
\end{itemize}   

\begin{figure}[ht]
\plotone{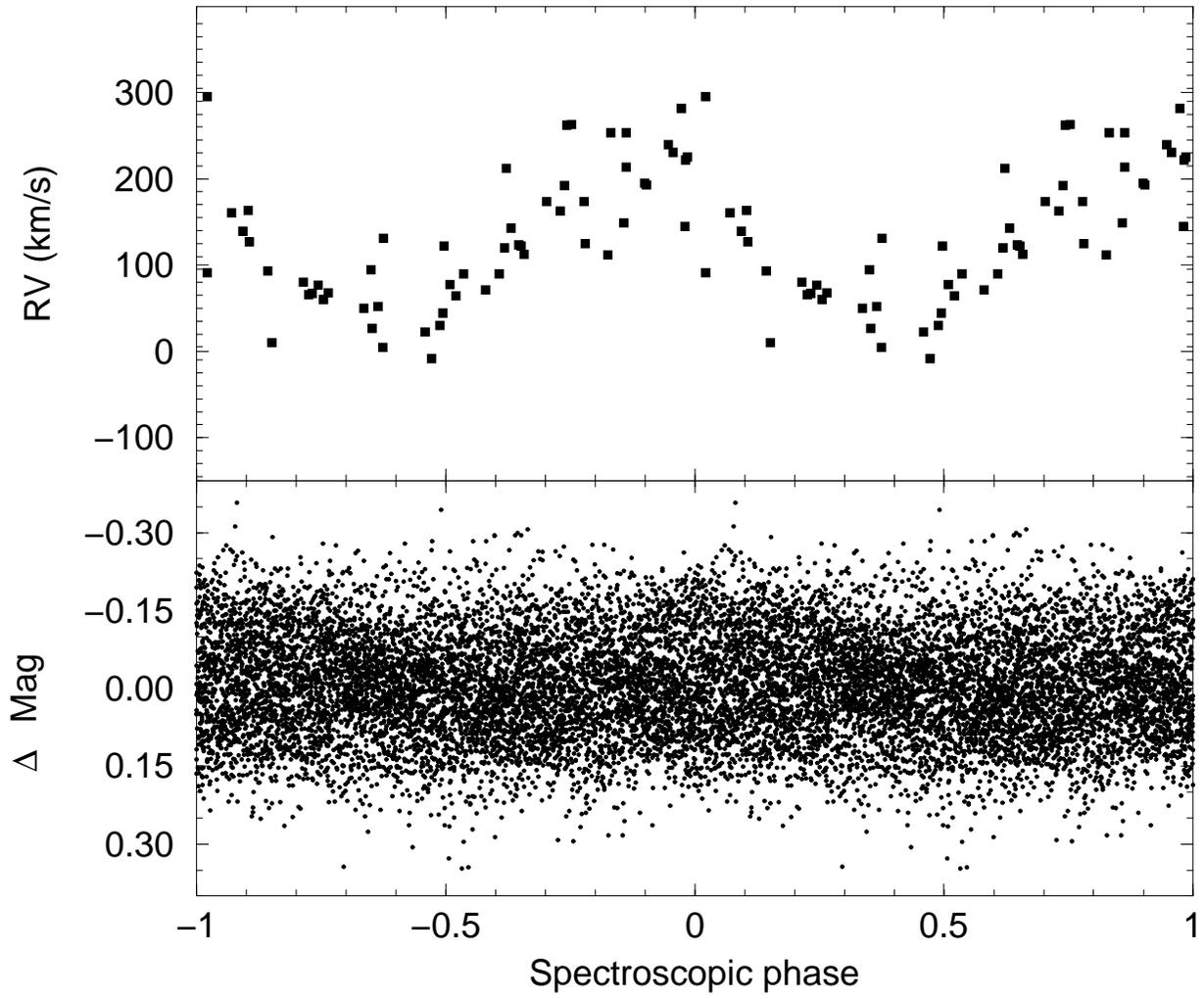}
\caption{The combined photometric data folded with the spectroscopic  
(orbital) period.
 }
\label{rvlc}
\end{figure}

 \noindent {\sl  The  photometric  period  exceeds  the  spectroscopic  by  2.4
times!}  This is an unprecedented case  for  a low  mass binary  system
unambiguously identified as a  Cataclysmic Variable.  In order to show
that the two periods do  not match we plotted the combined photometric
data folded with the spectroscopic  (orbital) period (Figure \ref{rvlc}), along
with the radial velocity curve obtained from measurements of the H$_\beta$
line from our observations. The  zero point for spectroscopic as well
as photometric phase, the $+/-$ crossing of the radial velocity, 
was taken from N02.

\section{Possible causes of the large photometric period}

\begin{enumerate}
\item The possibility of  rotational effects in the reference frame of  
the secondary star as well as a
triple/multiple component  system  can  be  effectively
ruled out based on the canonical  sinusoidal shape of the light curves,
the equality of the  amplitude of variation in the three  bands and the
absence of any spectral evidence of such kind.

\item  
  As a  dwarf nova with an orbital  period below the period gap,  FS Aur was
expected to be  a SU UMa type object and show  superhumps due to an eccentric 
disc (see 
the  extensive discussion  in the  recent paper  by  \cite{Patt01} and
references  therein).  The  number  of superhumpers  is  statistically
significant, theory and models of superhumps are well developed. Based
on  our  knowledge of  this  phenomenon  one  would expect  to  detect
superhumps  with   periods  exceeding  the  orbital  only   by  a  few
percent. Noting  that permanent superhumps were  detected in novalike
systems  with short  periods, a  dwarf nova  of SU  UMa  type produces
superhumps  only during superoutbursts.  FS Aur  in turn  never showed
superhumps  during decades of optical monitoring.  We  think that it  is 
extremely unlikely that the photometric period in  FS Aur is due to 
an eccentric
disc precession because the period difference is much too large.

\item The long  period could be the precession period of  a warped disc. One
possible model  to explain  the difference between  spectral (orbital)
period and the  photometric variations in FS Aur is  the presence of a
tilted accretion  disc which is  rigidly precessing with  the observed
photometric  period.  Similar models  have  been  proposed to  explain
long-term variations  in X-ray  binaries. The analytical  estimates of
the  induced   precession  period   are  given  by   \cite{Pap95}  and
\cite{Lar98}.   The ratio between  the binary  orbital period  and the
forced precession period is
 
\begin{equation}
\frac{P}{P_p} =\frac{3}{7}\ q\ \biggl(\frac{1}{1+q}\biggr)^{1/2}\biggl(\frac{R_o}{a}\biggr)^{3/2}\cos\,\delta
\end{equation} 

where $ q$  =  M${_\mathrm   s}$/M${_\mathrm  p}$  is  the  mass  ratio,
$R{_\mathrm  o}$  is  the  outer  radius  of  the  disc,  the  orbital
separation  is  $a$, and  $\delta$  is  the  orbital inclination  with
respect to  the disc.  For  mass ratios $q=0.2-20$, Larwood  found that
period ratios for precessing tilted  discs is in the range $P{_\mathrm
p}/P \approx  18-40$. If  the mean \cite{Pap77}  disc truncation
radius applies, then the lower value is $\approx10$.

   From the  absence of superoutbursts in  FS Aur we are  not bound to
the 3:1 resonance criterium, so the mass ratio can be different to the
$0.05 <q<0.25$  requirement for the eccentric  disc resonance. However,
it is hard to find reasonable values of masses in order to reach a rigid
precession of a gaseous Keplerian disc in  a system similar  to FS Aur.
The  secondary in  the  system should  be  late M  dwarf according  to
the relation between orbital period and spectral class of the secondary.
The absence   of  any
contribution in the  near infra-red  spectrum from the  secondary confirms
that it  is at  least later than  M\,0 with  M$_2 <$ 0.5\msun.  On the
other hand,  rough estimates of  M$_1$ yield values  somewhere between
0.34$<M_1<0.46$ (N02).  Thus, the  mass ratio $q$  can reach  at maximum
1.5, if we assume an M0 secondary with 0.5 M$_\sun$ and the lower limit on
M$_1$.  With  these parameters $P_{\mathrm p}/P_{\mathrm  orb} > 14$
according to \cite{Lar98} and the precession period in FS Aur should have
been of the order of a day ($\geq 20$ hours)

Radiation  driven  warping  of  accretion discs  \citep{Ogl01}  imposes
additional restrictions  on the parameters of binary  systems in order
to maintain stable precessing discs.

Thus, from a purely theoretical point of  view it is  very difficult to
justify a disc precession model.  From the observational point of view
the  contribution of the  disc in  emission is  inferior to  the spot,
whatever the  origin of the spot  is. Thus,  the spectroscopic data
do not provide strong evidence  of a luminous disc, whose precession
could lead to the observed photometric variations.

\item Could it be due to the rotation of the magnetic poles of the white dwarf?
 The colors and shapes of the optical light curve strongly support the
magnetic spin  version of the  photometric period; there  are numerous
similar light curves  in the literature on intermediate polars (IPs).  
For example we refer
to  \cite{Nor02}  and   references  therein.  The  intermediate  polar
scenario is so far the most  plausible way to immerse unrelated to the
orbital period in a close binary system.

With this interpretation, we run  basically into one important issue: 
why  is the period of
rotation  so slow? None of  the dozens of  known intermediate polars
show white dwarf spin  periods less than the orbital one. 
Yet,  there are no theoretical
restrictions on the  velocity of rotation of the  white dwarf in CVs. 
Generally,
one would expect the rotation periods of white dwarfs in  binary systems not to
be  very  different  from   single  WDs.  According  to  \cite{Spr01},
the distribution of rotational velocities of WDs permits periods as
the one observed in FS Aur.

However, the presence of a  magnetic field  anchored  on a  WD makes  almost 
impossible the existance of a stable spin  periods  less  than  0.68 P$_{\rm  orb}$
according to \cite{Kin99}  unless   the  magnetic  field  is   high  enough, so that the
magnetosphere of the WD overfills its  respective Roche lobe and locks the
system  into synchronous  rotation.  \cite{Som01}  considered  the spin
periods  of  magnetic  CVs  and   show  that  there  are  very  strong
restrictions on the spin periods in magnetic CVs.   
But we should  note here that
there are a few examples  of asyncroneous polars, that are supposed to
be jolted out  of synchronism by a nova explosion.  The period difference
is  again not  so big  as in  the case of FS Aur,  and any system would
 rapidly  tend to
establish synchronization, but little is known what happens during such
cataclysms and  how far it may  take the spin period.  Therefore, it is
difficult  to directly apply the intermediate polar  scenario  to the 
case  of FS  Aur in this pretext.

An assumption alternative  to the  longer  than  orbital spin  period  
might be  the possibility
that  the real white dwarf spin  period is of the order of 50--100  sec and
FS\,Aur is  a fast rotator as  AE\,Aqr and DQ\,Her. If  we assume also
that  the white  dwarf  in  FS\,Aur freely  precesses  then such  fast
rotation will result in the wobbling  of the X-ray projector together with the
WD. In a certain orientation of the magnetic pole angle to the orbital plane
(colatitude $\beta$),  the X-ray beam may  or may not  sweep across the
disk or its inner edge. What,  or how much that fan beam perpendicular
to the magnetic field lines  will illuminate may explain the nature of
this so far mysteriuos photometric  period. The idea of free precession
of WD was explored by \cite{Lei92}
%  Leins etal (1992 AA 261, 658).
They mention intermadiate polars as  a possible laboratory to test it.
According  to  \cite{Lei92},  Euler's  frequency  of free  precession
$\omega_e$ of a rigid and axially symmetric body will be:

\begin{equation}
\omega_e = \alpha(\rho_c)\Omega^3
\end{equation}

In case of an elastic body the Chandler frequency $\omega_c$ is used instead:

\begin{equation}
\omega_c = \beta(\rho_c)\Omega^3
\end{equation} 

where  $\alpha_c(\rho_c)$ and  $\beta_c(\rho)$ are  coefficients that
are determined by the central density  $\rho_c$ or a mass M, radius R,
and moment of inertia A of  the white dwarf.  The results of numerical
calculations for the  realistic models of a white  dwarf with different
central densities are presented  in the Table\,2 of \cite{Lei92}.  For
a mass of  a white dwarf  between 0.55-0.63  M$_\sun$, 
$\alpha_c(\rho_c)$   ranges    between   2.109   and    1.299,   while
$\beta_c(\rho)$ varies  from 0.219  to 0.124, correspondingly.   Thus, a
white dwarf rotating with $50 \leq$ P$_{spin} \leq 100$ sec would have
a free  precession period of P$_{prec}=205$ min  depending on which model
(Euler or Chandler) we adopt.

   A  good  test for this intermediate  polar scenario is the  
X-ray  light curve.  The spectral hardness  ratio is  in agreement  with X-ray  spectra from
other known  IPs. We  have folded the X-ray data of the ROSAT pointed
observation
with the photometric  and spectroscopic  period, respectively. 
The corresponding  curves are
shown in  Figure \ref{xfold}. None  of the two figures  contradicts the
presence  of a periodic signal  with the corresponding  period in  the light
curve.
FS\,Aur was observed 5 times
with 10-20 min each, at time intervals of 3-4 hrs. The periodic pattern in both 
panels most probably is the observing window.  This is what causes
the left panel in the paper to show the "coherent" pattern - it is just
the observing window pattern. The statistics are very poor and the limit on the pulse
fraction is only $<$80\%. Therefore, depending on the binning of a handful
of photons one can reach misleading results.   
New X-ray observations may are needed and may  help to sort this out.

\end{enumerate}

\section{Conclusions}

There were arguments  in the literature as to whether  SW Sex objects owe
their unusual characteristics to  the phenomenon known as intermediate
polars \citep{War95}.   V795 Her has been proposed  to be intermediate
polar and SW Sex object at the same time (see \cite{Cas96}). The newly
discovered KUV 03580+0614 \citep{Szk01} is a good candidate to qualify
for  both categories.  There  are other,  less obvious  examples.  These
objects lack  the detection of a stable, clearly  detectable spin period.
In  FS Aur,  we were  able to  observe two  distinct,  unrelated periods
similar to  the intermediate polars. There are  very firm restrictions
on  upper limits  of the spin  periods  of IPs,  and 205  min is  certainly
excluded for  FS\,Aur as the period of  rotation of the  WD. Therefore, we
propose  that the  WD in  FS\,Aur is  a fast  rotator  with P$_{spin}$
around 50 to  100 sec and freely precesses  with P$_{prec}=205$ min. The
cause of optical modulations with spin and side-band periods in IPs is
still a matter  of debate, but regardless of  its nature the amplitude
of the modulation is simply a geometrical effect of the orientation of the
X-ray  beam  relative to  the line of sight and/or  orbital  plane.
Therefore, it may also explain why the observed phenomenon is unique so
far.  The number of IPs  is still statistically small and the detection of
this effect might be merely a matter of a distinct  magnetic pole colatitude.
We  suggest that this phenomenological  model 
can be observationally tested with fast optical  photometry  and/or  
more sensitive  X-ray observations (Chandra/XMM).

While  the combination  of the SW  Sex  and intermediate  polar models  is
appealing, there are big difficulties  in squeezing FS Aur into each of
these.  Thus,  FS Aur may  be a rare,  new kind of  cataclysmic variable
which can be defined as a  dwarf nova, which in addition to its orbital period
shows a 2-3 times longer photometric period in its light curve.
% and still needs an explanation of its extraordinary behavior. 
Recently \cite{WW}
discoverd another similar system. GW Lib is another short period CV, that apart 
from its 1.28 hour spectroscopic period shows well defined photometric modulation
of light with 2.09 hour periodicity. As in the case with FS Aur, the reasons for such 
behaviour are not yet clear. However we can talk of a new phenomenon within 
Cataclysmic Variables.  FS\,Aur and GW\,Lib can be  
classified as new type of CVs, which still needs explanation of
its nature.    

\acknowledgments

%%%J
%We are grateful to Juan Pablo II

This work was supported in part by CONACYT projects 25454-E, 36585-E.

%\clearpage

%\subsection{Mathematical Symbols}
%
%The following  macros produce various troublesome or laborious mathematical symbols.
%\begin{center}
%\begin{tabular}{llll}
%\verb"\deg" & (\deg)&\verb"\farcs"&(\farcs)\\
%\verb"\sun"& (\sun)&\verb"\fp"&(\fp)\\
%\verb"\earth"&(\earth)&\verb"\micron"&(\micron)\\
%\verb"\la"&({$\la$})&\verb"\onehalf"   &(\onehalf)\\
%\verb"\ga"&({$\ga$})&\verb"\onethird"&(\onethird)\\
%\verb"\arcmin"&(\arcmin)&\verb"\twothirds"&(\twothirds)\\
%\verb"\arcsec"&(\arcsec)&\verb"\onequarter"&(\onequarter)\\
%\verb"\fd"&(\fd)&\verb"&\threequarters"&(\threequarters)\\
%\verb"\fh"&(\fh)&\verb"\ubvr " &(\ubvr)\\
%\verb"\fm"&(\fm)&\verb"\ub"    &(\ub)\\
%\verb"\fs"&(\fs)&\verb"\bv"    &(\bv)\\
%\verb"\fdg"&(\fdg)&\verb"\vr"  &(\vr)\\
%\verb"\farcm"&(\farcm)&\verb"\ur"      &(\ur)\\
%\end{tabular}
%\end{center}
%
%\bigskip

\clearpage
\begin{table}[t]
\caption{Log of optical observations of FS Aur.}
\begin{tabular}{ccccc}\hline \hline
 HJD start  & Duration  &  Time of exposure &  Band     & Telescope  \\ 
 2450000+   &  min      &     sec.          &           &            \\ \hline
 1165.72    &  193      &      60           &   white   &   CBA          \\
 1170.59    &  350      &      60           &   white   &   CBA          \\
 1174.51    &  82       &      60           &   white   &   CBA          \\
 1178.50    &  222      &      60           &   white   &   CBA          \\
 1179.50    &  378	&      60	    &	white	&   CBA           \\ 
 1182.47    &  476	&      60	    &	white	&   CBA            \\
 1184.57    &  313	&      60	    &	white	&   CBA           \\ 
 1185.64    &  482	&      60	    &	R	&   CBA            \\ 
 1186.64    &  506	&      60	    &	R	&   CBA             \\
 1187.74    &  345	&      60	    &	R	&   CBA        \\      
 1188.58    &  472	&      60	    &	R	&   CBA	        \\     
 1194.52    &  304	&	60	    &	white	&   CBA	         \\    
 1510.68    &  283      &	60	    &   white	&   CBA       \\
 1512.58    &  425	&	60	    &	white	&   CBA	         \\ 
 1513.59    &  430	&	60	    &	white	&   CBA	          \\
 1563.62    &  113	&      60	    &	white	&       CBA         \\
 1571.60    &  123      &      15	    &	    V	&   1.3m/MDM-CBA  \\ 
 1575.62    &  417	&      60	    &	white	&	CBA	   \\
 1584.60    &  49       &      15	    &	   bg38 &   2.4m/MDM-CBA \\ 
 1585.40    &  238      &      80	    &	   white&  0.35m/CBA	 \\    
 1586.24    &  94       &      80	    &	   white&  0.35m/CBA	    \\
 1619.68    &  116      &     180	    &	    R	&   1.5m/SPM     \\ \hline
\end{tabular}
\label{tab1}
\end{table}

\clearpage
\begin{tabular}{ccccc}\hline\hline
 HJD start  & Duration  &  Time of exposure &  Band     & Telescope  \\ 
 2450000+   &  min      &     sec.          &           &            \\ \hline
 1621.65    &  164      &     100	    &	    R	&   1.5m/SPM	  \\ 
 1622.64    &  184      &     180	    &	    V	&   1.5m/SPM	   \\
 1864.84    &  322      &     150	    &	    V	&   1.5m/SPM	  \\ 
 1865.80    &  298      &     150	    &	    V	&   1.5m/SPM	   \\ 
 1928.63    &  409      &     180	    &	    B	&   1.5m/SPM	    \\
 1941.63    &  368      &     600           &  3800-5200\AA & 2.1m/SPM     \\
 1942.62    &  299      &     600           &  3800-5200\AA & 2.1m/SPM     \\
 1946.65    &  315      &      30,20	    &	    B,R &   1.5m/SPM       \\ 
 2206.84    &  289      &	30	    &	    B	&   1.5m/SPM      \\
 2259.78    &  157      &	30	    &	    B	&   1.5m/SPM   \\
 2260.76    &  178      &	30	    &	    B	&   1.5m/SPM  \\ \hline
\end{tabular}

\begin{tabular}{l}
SPM - San Piedro Martir Observatory, Mexico \\
MDM - MDM Observatory \\
CBA - Center of Backyard Astronomy  \\
\end{tabular}

\clearpage
\clearpage

%\begin{figure}[ht]
%\plotone{tovmassian_fig2.ps}
%\caption{The  power spectrum  of  the  H$_\beta$ radial  velocity 
%measurements from both data sets.  The
%overplotted thick curve  is the CLEANed power of  the same data. There
%are  two  close   peaks  at  the  orbital  period   with  the  highest
%corresponding  to  85.65 (0\fd059479)  minutes  and  smaller at  85.72
%(0\fd059528) minutes  with only  4 sec (0\fd00005)  difference between
%them.   We adopted  the former  as  the measure  of the  spectroscopic
%(orbital) period  of the system.
%}
%\label{powsp}
%\end{figure}
%\clearpage

\clearpage

\clearpage

\clearpage
\clearpage


\clearpage

\clearpage

\clearpage

\clearpage
\begin{figure*}[ht]
%%\begin{picture}
\plottwo{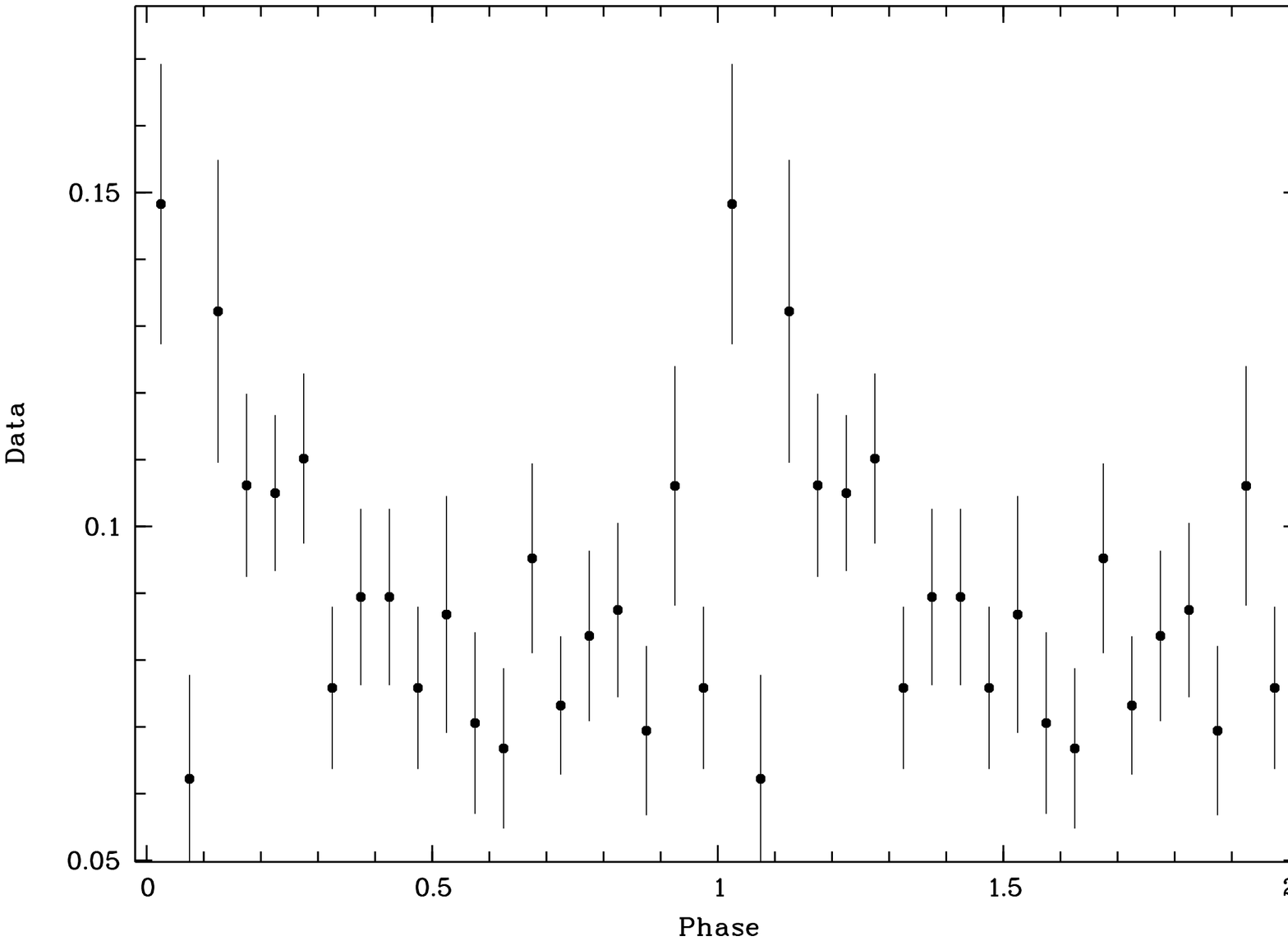}{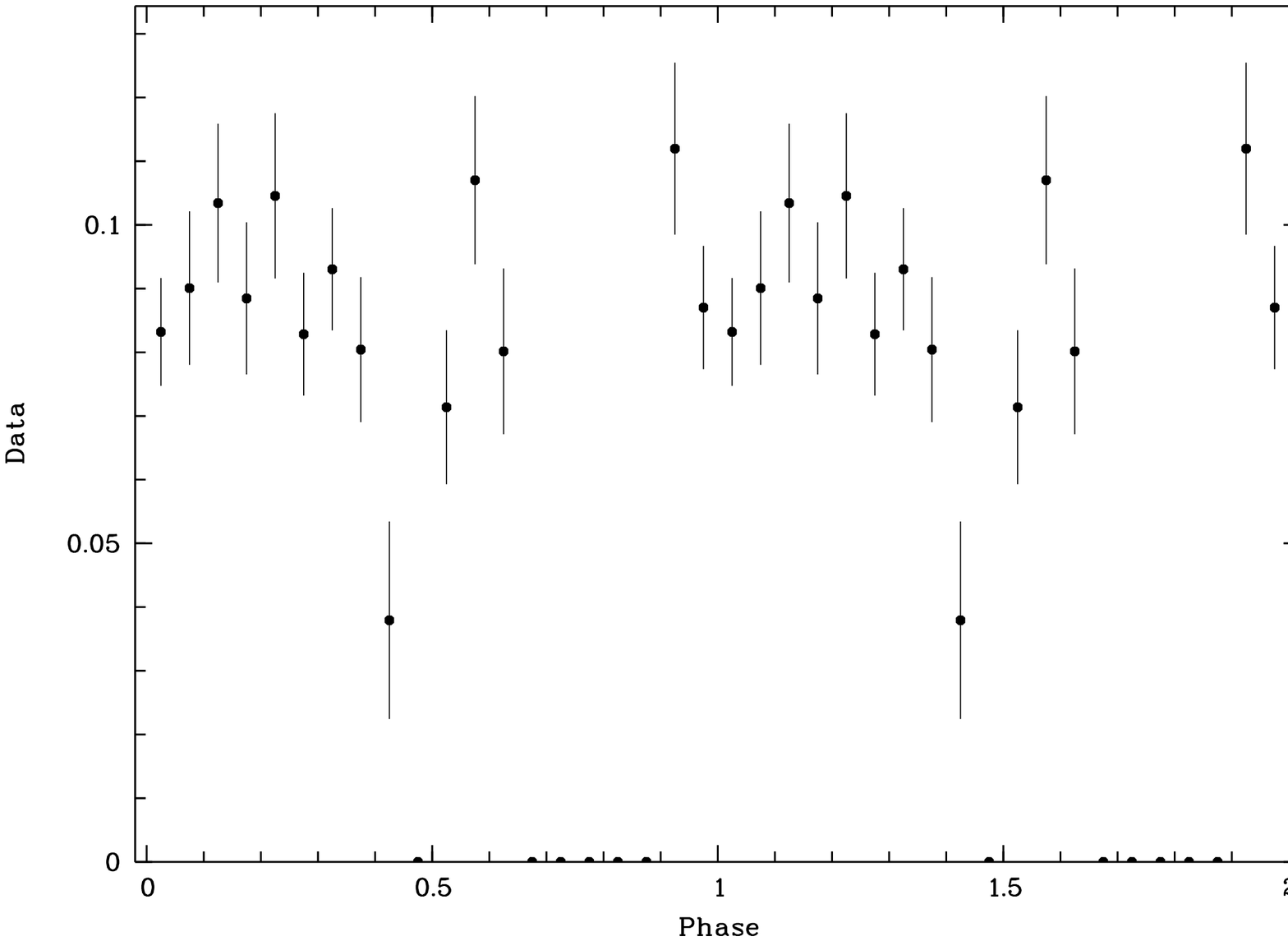}
\caption{Folded X-ray light curves of FS Aur, obtained from the
  ROSAT PSPC observation of 23 September 1993, and folded with the
  spectroscopic (i.e. orbital; left panel) and photometric (205min) period (right
  panel). 
 }
\label{xfold}
\end{figure*} 

\end{document}